\documentclass{article}
\usepackage{amsfonts,amssymb, amsmath}
\usepackage[english]{babel}
\usepackage{graphicx}

\def\nn{\nonumber }
\def\bq{\begin{equation}}
\def\eq{\end{equation}}
\def\ben{\begin{eqnarray}}
\def\en{\end{eqnarray}}

\newtheorem{prop}{Proposition}
\newtheorem{defi}{Definition}

\begin{document}

\title{B\"{a}cklund transformations and  divisor doubling}
\author{A.V. Tsiganov \\
\it\small St.Petersburg State University, St.Petersburg, Russia, Udmurt State University,  Izhevsk,  Russia\\
\it\small e-mail:  andrey.tsiganov@gmail.com}

\date{}
\maketitle

\begin{abstract}
 In classical mechanics well-known cryptographic algorithms and protocols can be very useful for construction canonical transformations preserving form of Hamiltonians.  We consider application of a standard generic divisor doubling for construction of new auto B\"{a}cklund transformations for the Lagrange top and H\'{e}non-Heiles system separable in parabolic coordinates.

 \end{abstract}

\section{Introduction}
\setcounter{equation}{0}
The method of additive separation of variables in Hamilton-Jacobi equation, at least, in its most elementary forms
such as separation of variables in  elliptic, parabolic, spheroconical etc.  coordinates, is a very important tool in analytical mechanics.
Following to Jacobi works, an $n$-tuple $(H_1,\ldots,H_n)$ of functionally independent Hamiltonians in the involution will be said to be separable in a set of canonical coordinates $(x_1,\ldots,x_n,y_1,\ldots,y_n)$ if there are exist $n$ relations of the form
\[
\Phi_i(x_i,y_i,H_1,\ldots,H_n)=0,\qquad i=1,\ldots,n,\qquad \mbox{with}\quad \mbox{det}\left[\frac{\partial \Phi_i}{\partial H_j}\right]\neq 0\,.
\]
The reason for this definition is that the stationary Hamilton-Jacobi equations   $H_1=\alpha_1,\ldots,H_n=\alpha_n$ can be collectively solved by the additively separated complete integral
\[
S(x_1,\ldots,x_n;\alpha_1,\ldots,\alpha_n)=\sum_{i=1}^n S_i(x_i;\alpha_,\ldots,\alpha_n),
\]
where the $S_i$ are found by quadratures.

According to the Jacobi theorem if  we  substitute pairs of canonical coordinates $(x_i,y_i)$ into a system of separation relations \[\Phi_i(x,y,\alpha_1,\ldots,\alpha_n)=0, \qquad i=1,\ldots,n,\qquad \mbox{with}\quad \mbox{det}\left[\frac{\partial \Phi_i}{\partial \alpha_j}\right]\neq 0\,.
\]
and solve these equations with respect to $\alpha_1,\ldots,\alpha_n$, we  obtain functionally independent Hamiltonians in the involution. Each separation relation $\Phi_i (x,y,\alpha_1,\ldots,\alpha_n)$ defines the curve $X_i$ on a $(x,y)$-plane, that allows us to consider integrable system $\{H_1,\ldots,H_n\}$ as an ordered set of points $\{P_1,\ldots,P_n\}$ on an ordered product $X_i\times\cdots\times X_n$.

For the systems separable in elliptic, parabolic, spheroconical etc.  coordinate systems  we have $n$ equivalent separation relations \[\Phi_i(x,y,\alpha_1,\ldots,\alpha_n)=\Phi_j(x,y,\alpha_1,\ldots,\alpha_n)\,\qquad i,j=1,\ldots,n.\]
If the corresponding  plane curve $X$ has a necessary properties, we can consider formal sum of  prime divisors $D=\sum_{i=1}^n P_i$ instead of an ordered set of points and  identify this formal sum  with a $n$-degree divisor $D$ on $X$ after compactification.

In hyperelliptic curve cryptography divisor $D$  on a hyperelliptic curve $X$ of genus $g$ plays the role of a message, which can be coding to a cryptogram $D''$, which is another divisor, using some standard cryptographic protocol. We can use secret protocols developed for full-degree divisors $n=g$ or degenerate divisors $n\neq g$, protocols based on divisor arithmetic or post quantum protocols based on isogenies etc, see  \cite{cant87,cost12,cost16, hand06,kat,suth16} and references within. Our main goal is to understand what means standard coding/decoding operations in classical mechanics and how to use these operations and the corresponding cryptograms in the theory of finite-dimensional integrable systems.

In this note we  apply standard coding  operations associated with addition and  scalar multiplication of divisors
\[D''=D+D'\,\qquad D''=[2]D\]
to the divisors associated with Lagrange top and H\'{e}non-Heiles system separable in parabolic coordinates.
In cryptography divisor $D'$ is a secret key, whereas scalar multiplication is a standard part of the  keyless secret systems.  We prove that in classical mechanics the corresponding coding operations are canonical transformations of valence one or  two, which preserve  the form of Hamilton-Jacobi equations.   In   \cite{kuz02,kuz04}  reader could find another explicit formulae for canonical transformations for the Lagrange top and  H\'{e}non-Heiles system, which preserve the form of Hamiltonians,  that determined our choice of examples.
Canonical transformations preserving form of Hamiltonians we will call auto-B\"{a}cklund transformations following to Toda and Wadati \cite{toda75}. There are also many other definitions of the auto-B\"{a}cklund transformations for Hamilton-Jacobi equations
\cite{fed00,kuz99,kuz02,skl00},  but we prefer to use the oldest one.

 \section{Divisor arithmetic on hyperelliptic curves}
 Let us  reproduce some definitions and facts from the following textbooks  \cite{eh16,har77}.

A prime divisor on a smooth  variety $X$ over a field $k$ is an
irreducible closed subvariety $Z\subset X$ of codimension one, also defined over $k$.
\begin{defi} A divisor is a finite formal linear combination
\[D =\sum_i m_iZ_i,\qquad m_i\in \mathbb Z\,, \]
of prime divisors. The group of divisors on $X$, which is the free group on the prime divisors, is denoted Div$X$.
\end{defi}
The group of divisors Div$X$ is an additive abelian group under the formal addition rule
 \[\sum m_i Z_i+\sum n_i Z_i=\sum (m_i+n_i) Z_i\,.\]
To define  an equivalence relation on divisors we  use the rational functions on $X$.
Function  $f$ is a quotient of two polynomials; they are
each zero only on a finite closed subset of codimension one in $X$, which is therefore the union of finitely many prime divisors. The difference of these two subsets  define  a principal divisor $div f$
 associated with function $f$. The subgroup of Div$X$ consisting of the principal divisors is denoted by Prin$X$.
\begin{defi}
Two divisors $D, D'\in \mbox{Div} X$ are linearly equivalent
\[D\approx D'\]
if their difference $D-D'$ is principal divisor
\[
D-D'=div(f)\equiv 0\quad \mathrm{mod\, Prin}X\,.
\]
\end{defi}
The Picard group of $X$ is the quotient group
\[
\mbox{Pic}X =\dfrac{\mbox{Div}X}{\mbox{Prin}X}=\dfrac{\mbox{Divisors defined over k}}{\mbox{Divisors of functions defined over k}}\,.
\]
For a general (not necessarily smooth) variety X, what we have defined is not the Picard group, but the Weil divisor class group. For an irreducible normal variety $X$, the Picard group is isomorphic to the group of Cartier divisors modulo linear equivalence.

The Picard group is a group of  divisors modulo principal divisors, and the group operation is formal addition modulo the equivalence relations. These group operations define so-called  arithmetic of divisors in Picard group
\bq\label{add-jac}
D+ D'=D''\qquad\mbox{and}\qquad [\ell] D=D''\,,
\eq
where $D,D'$ and $D''$ are divisors, $+$ and $[\ell]$ denote addition and scalar multiplication by an integer, respectively.

Let $X$ be a hyperelliptic curve of genus $g$  defined by equation
\bq\label{h-curve}
 y^2 + h(x)y = f(x),
 \eq
where $f(x)$ is a monic polynomial of degree $2g + 2$ with distinct roots, $h(x)$
is a polynomial with deg$h\leq g$.  Prime divisors are rational point on $X$  denoted $P_i = (x_i, y_i)$,  and $P_\infty$ is a point at infinity.
\begin{defi}
Divisor $D = \sum m_iP_i$, $m_i\in \mathbb Z$ is a formal sum of points on the curve, and degree of  divisor $D$ is the sum $\sum m_i$ of the multiplicities of points in  support of the divisor
\[
\mbox{supp}\left(\sum m_iP_i\right)=\bigcup_{m_i\neq 0} P_i\,.
\]
\end{defi}
 Quotient group of $\mbox{Div}X$ by the group of principal divisors Prin$X$ is called the divisor class group or Picard group.
 Restricting to degree zero, we can also define $\mbox{Pic}^0 X= \mbox{Div}^0 X/\mbox{Prin}X$. The groups $\mbox{Pic}X$ and $\mbox{Pic}^0 X$ carry essentially the same information on $X$, since we always have
 \[
\mbox{Pic}X/\mbox{Pic}^0 X \cong \mbox{Div}X/\mbox{Div}^0 X\cong\mathbb Z\,.
 \]
 The divisor class group, where the elements are equivalence classes of degree zero divisors on $X$,  is isomorphic to the Jacobian of  $X$. By abuse of notation, a divisor and its class in Pic$X$ will usually be denoted by the same symbol.

In order to describe equivalence classes we can use semi-reduced and reduced divisors.
\begin{defi}
A  semi-reduced divisor is divisor of the form
\[
D=\sum m_i P_i -\left( \sum m_i \right) P_\infty,,
\]
where $m_i>0$, $P_i\neq -P_j$ for $i\neq j$, no $P_i$ satisfying  $P_i=-P_ i$ appears more than once.
\end{defi}
Because semi-reduced divisors are not unique in their equivalence class we introduce reduced divisors.
\begin{defi}
A semi-reduced divisor $D$ is called reduced if  $\sum m_i\leq g$, i.e. if the sum of multiplicities is no more that genus of curve $C$.  The reduced degree  or weight of reduced divisor $D$ is defined as $w(D)=\sum m_i$.
\end{defi}
This is a consequence of Riemann-Roch theorem for hyperelliptic curves that for each divisor $\tilde{D}\in \mbox{Div}^0X$ there is a unique reduced divisor $D$ so that $D\approx \tilde{D}$. For a thorough treatment see \cite{eh16,har77}.

Using  reduced divisors $D$ instead their equivalence classes we can describe fast and efficient algorithms for arithmetic on hyperelliptic curves. In  \cite{mum} Mumford found polynomial representation of group elements $D=(u(x),v(x))$
\[u(x)=\prod(x-x_i)^{m_i}\,,\quad v(x_i)=y_i\,,\quad \mbox{deg}(V)<\mbox{deg}(U)\leq g\,,\quad v^2-f\equiv 0\,\mbox{mod}\,u\,.
\]
Here monic polynomial $u(x)$ may have multiple roots  and $v(x)$ is tangent to the curve according to multiplicity roots.  In fact these  polynomials were introduced by Jacobi in the framework of the classical mechanics, see historical remarks, discussion and  modern  applications of these Jacobi polynomials in  \cite{in07,van01}.

In \cite{cant87} Cantor proposed the following algorithm for performing arithmetic computations   in Picard group of hyperelliptic curves $X$ defined by equation  (\ref{h-curve}):
\[
\begin{array}{l}
$------------------------------------------------------------------------------------------------------------$\\
\mbox{\textbf{Input}}\quad  D=(u_1,v_1),\qquad D'=(u_2,v_2);\qquad
\mbox{\textbf{Output}}\quad  D''=(u_3,v_3)=D+D'\\
$------------------------------------------------------------------------------------------------------------$\\
1.\quad d=\mbox{gcd}(u_1,u_2,v_1+v_2+h)=S_1u_1+S_2u_2+S_3(v_1+v_2+h)\\
\\
2.\quad U\leftarrow\frac{u_1u_2'}{d^2},\qquad V\leftarrow\frac{S_1u_1v_2+S_2u_2v_1+S_3(v_1v_2+f)}{d}\,\mbox{ mod }\, U\\
\\
3.\quad \mbox{while deg}(U)>g \\
\hskip4truecm   U'\leftarrow\frac{f-hV-V^2}{U},\quad V'\leftarrow-h-V\mbox{ mod } U'\\
  \hskip4truecm U\leftarrow\mbox{MakeMonic}(U'),\qquad V\leftarrow V'\mbox{ mod } U\\
  \\
4.\quad u_3\leftarrow U\,,\qquad v_3\leftarrow V \\
5.\quad \mbox{return}\,(u_3,v_3)\\
$------------------------------------------------------------------------------------------------------------$
\end{array}
\]
 This algorithm  consists of two stages: the composition stage, based on Gauss's classical composition of binary quadratic forms, which generally outputs an unreduced divisor with coordinates $(U,V)$, and  the reduction stage, which transforms the unreduced divisor
into the unique reduced divisor.  The Cantor algorithm is quite slow due to its versatility and now we have a lot of other algorithms and their professional computer implementations for the divisor arithmetics.

\subsection{Arithmetic on genus 2 hyperelliptic curves}
For $g=2$ Cantor’s algorithm has since been substantially optimized in work
initiated by Harley \cite{har00} , who was the first to obtain practical explicit formulas in genus two, and
extended by Lange \cite{lange}, who, among several others, generalized and significantly
improved Harley’s original approach, see dicsussion in \cite{cost12}.

Let us present the Harley algorithm for addition and doubling:\\
\begin{minipage}[h]{0.49\linewidth}
\[
\begin{array}{l}
$-------------------------------------------------$\\
\mbox{\textbf{Input}}\quad  D=(u_1,v_1),\quad D'=(u_2,v_2);\\
\mbox{\textbf{Output}}\quad  D''=(u_3,v_3)=D+D'\\
$-------------------------------------------------$\\
1.\quad U\leftarrow u_1u_2\\
2.\quad S\leftarrow (v_2-v_1)/u_1\,\mbox{mod}\,u_2\\
3.\quad V\leftarrow Su_1+v_1\,\mbox{mod}\, U\\
4.\quad U\leftarrow (V^2+hV-f)/U\\
5.\quad \mbox{Make}\, U \,\mbox{monic}\\
6.\quad V\leftarrow V\,\mbox{mod}\,U\\
7.\quad u_3\leftarrow U,\,v_3\leftarrow-(V+h)\,\mbox{mod}\,u_3\\
8.\quad \mbox{return}\,(u_3,v_3)\\
$-------------------------------------------------$
\end{array}
\]
\end{minipage}
\begin{minipage}[h]{0.49\linewidth}
\[
\begin{array}{l}
$-------------------------------------------------$\\
\mbox{\textbf{Input}}\quad  D=(u_1,v_1),\quad \mbox{gcd}(u_1,h)=1;\\
\mbox{\textbf{Output}}\quad  D''=(u_3,v_3)=[2]D\\
$-------------------------------------------------$\\
1.\quad U\leftarrow u_1^2\\
2.\quad S\leftarrow (2v_1+h)^{-1}(f+hv_1-v_1^2)/u_1\,\mbox{mod}\,u_1\\
3.\quad V\leftarrow Su_1+v_1\,\mbox{mod}\, U\\
4.\quad U\leftarrow (V^2+hV-f)/U\\
5.\quad \mbox{Make}\, U \,\mbox{monic}\\
6.\quad V\leftarrow V\,\mbox{mod}\,U\\
7.\quad u_3\leftarrow U,\,v_3\leftarrow-(V+h)\,\mbox{mod}\,u_3\\
8.\quad \mbox{return}\,(u_3,v_3)\\
$-------------------------------------------------$
\end{array}
\]
\end{minipage}\\
Composition parts from Step 1 to step 3 are based on  the Chinese remainder
theorem for addition  and  Newton iterations for doubling, respectively.
We can take computer implementation of these algorithms from \cite{har00} and directly apply  computer programs to divisors associated with integrable systems.

Let us also present well-known explicit formulae for these coding operations in the simplest case of the  genus two hyperelliptic curve defined by equation (\ref{h-curve}) with $h(x)=0$
 \bq\label{eq-c}
 X:\quad y^2=f(x)\,,\qquad f(x)= a_5x^5+ a_4x^4+ a_3x^3+ a_2x^2+a_1x+a_0\,.
 \eq
Consider addition of  full degree reduced divisor (message)
\[
D: \quad\mbox{supp}(D)=\{ P_1, P_2 \}\cup \{ P_\infty \}\,,  w(D)=2
\]
with other reduced divisor $D'$ (secret key) in the following cases
\bq\label{cases}
\begin{array}{llll}
1.&\quad D+D'=D''\,,\qquad &w(D')=2\,,\quad&\mbox{supp}(D')=\{ P'_1,P'_2\}\cup \{P_\infty\}\,,\\
\\
2.&\quad [2]D=D''\,,\qquad& D'=D\,,\quad&\mbox{supp}(D')=\{ P_1,P_2\}\cup \{P_\infty\}\,,\\ \\
3.&\quad  D+D'=D''\,,\qquad& w(D')=1\,, \quad&\mbox{supp}(D')=\{ P'_1\}\cup \{P_\infty\}\,.
\end{array}
\eq
The result  is  full degree reduced divisor $D''$ (cryptogram) with  supp$(D'')=\{ P''_1,P''_2\}\cup \{P_\infty\}$ and $w(D'')=2$.

In the second case operation
\bq\label{doub-gen}
[2]D=D+D=D''
\eq
is called doubling of divisor $D$, i.e. scalar multiplication on integer $\ell=2$. Its inverse is called halving of $D''$ and  for a given $D''$ equation (\ref{doub-gen}) has $2^{2g}$ solutions, any two of which differ by a 2-torsion divisor \cite{ghar00},  for efficient implementation see \cite{half09} and references within. Below we do not consider halving,  tripling  and other operations because explicit formulae  for the corresponding canonical transformations are quite bulky and unreadable. The third case in (\ref{cases}) corresponds to one-point  auto B\"{a}cklund transformations from \cite{kuz02,skl00}.

Let us consider intersection of $X$  with the second plane curve defined by equation
\bq\label{eq-p}
y-\mathcal P(x)=0\,,\qquad \mathcal P(x)=b_3x^3+b_2x^2+b_1x+b_0\,.
\eq
in the framework of the standard  intersection theory \cite{ab,eh16,kl05}. Substituting $y=\mathcal P(x)$ into the equation  (\ref{eq-c}),
we obtain the so-called Abel polynomial \cite{ab}
 \[
\psi(x)=\mathcal P(x)^2-f(x)\,,
\]
 which has no multiple roots in the  first and third cases and has double roots in the second  case:
\[
\begin{array}{cl}
1.&\quad\psi(x)=b_3^2(x-x_1)(x-x_2)(x-x'_1)(x-x'_2)(x-x''_1)(x-x''_2)\,,\\
\\
2.&\quad \psi(x)=b_3^2(x-x_1)^2(x-x'_1)^2(x-x''_1)(x-x''_2)\,,\\
\\
3.&\quad \psi(x)=-  a_5(x-x_1)(x-x_2)(x-x'_1)(x-x''_1)(x-x''_2)\,,\quad b_3=0\,.\\
\end{array}
\]
The cases 1 and 2 are presented in   Figure 1, which is standard  Clebsch's geometric interpretation of the Abel results \cite{kl05}
\begin{figure}[!ht]
\begin{minipage}[h]{0.5\linewidth}
\center{\includegraphics[width=0.75\linewidth, height=0.3\textheight]{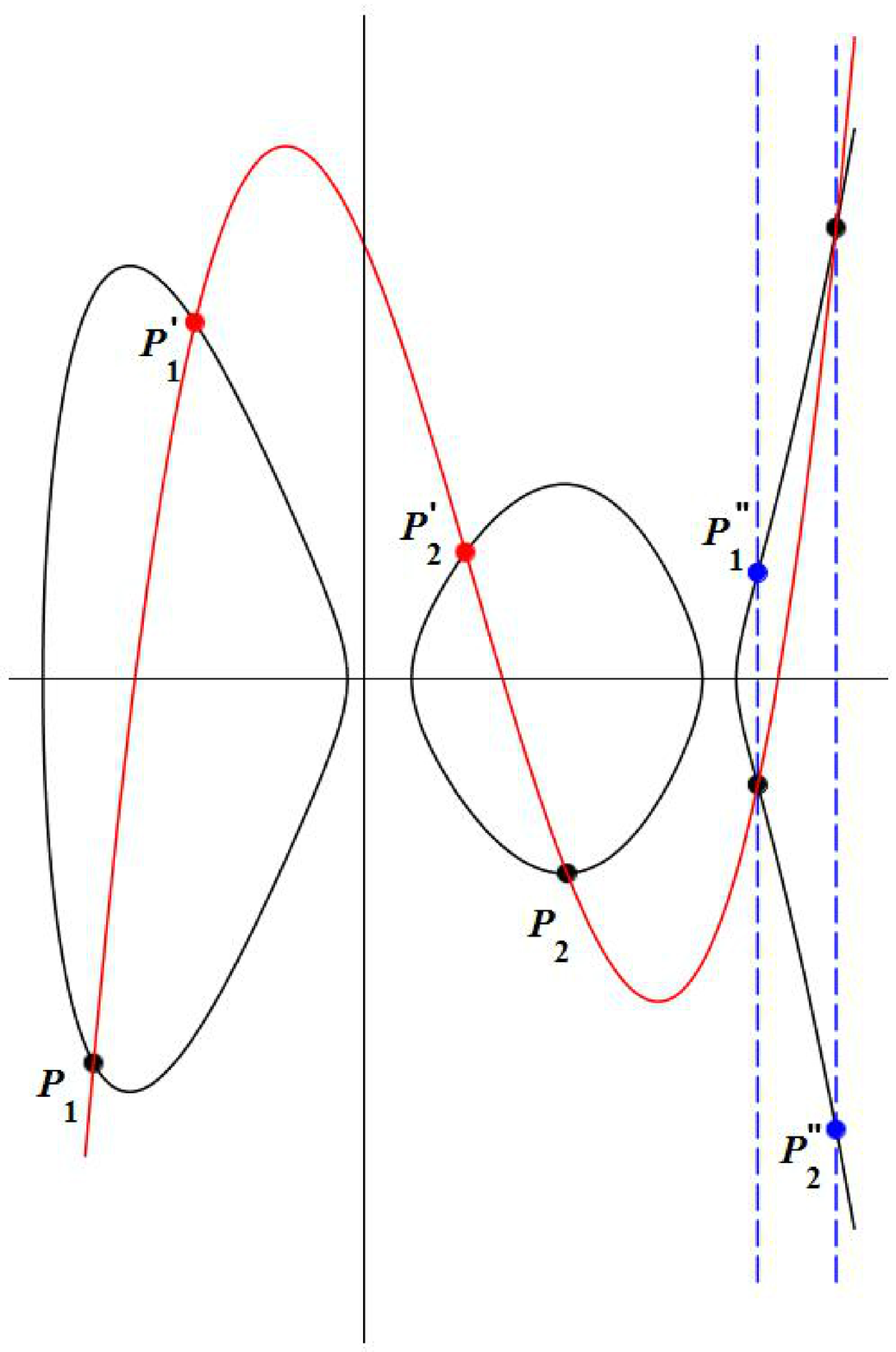} \\ Case 1: $(P_1+P_2)+(P'_1+P'_2)=P''_1+P''_2$}
\end{minipage}
\hfill
\begin{minipage}[h]{0.49\linewidth}
\center{\includegraphics[width=0.75\linewidth,height=0.3\textheight]{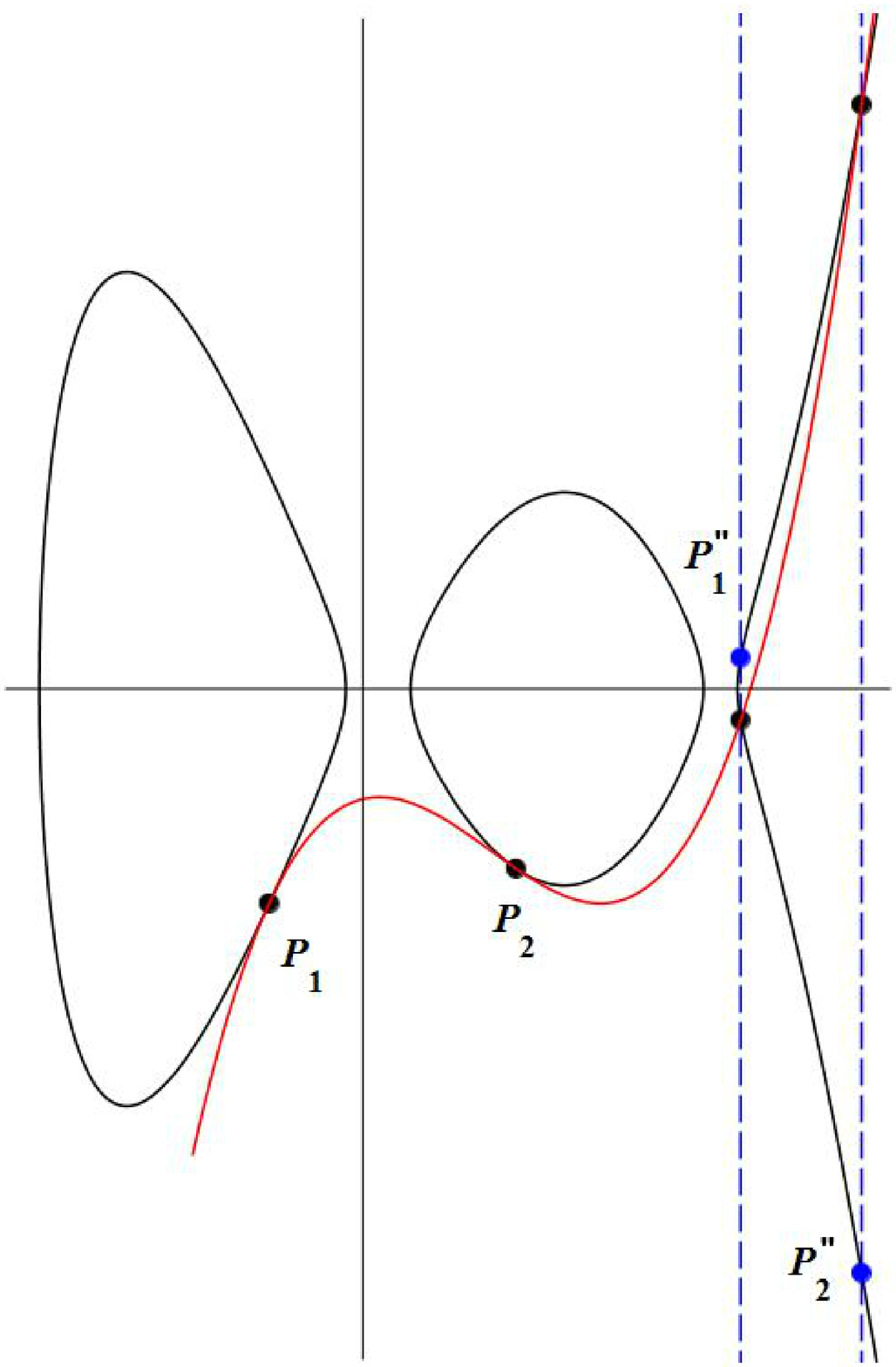} \\ Case 2: $[2](P_1+P_2)=P''_1+P''_2$}
\end{minipage}
\end{figure}

Equating coefficients of $\psi$  in the first case gives
{\setlength\arraycolsep{0pt}
\ben
&&x''_1+x''_2= -x_1-x_2-x'_1-x'_2+\frac{a_5-2b_2b_3}{b_3^2}\,, \label{g2-x12}\\
\nn\\
&&x''_1x''_2=\frac{2b_1b_3+b_2^2-a_4}{b_3^2}
-(x_1+x_2+x'_1+x'_2)(x''_1+x''_2)
-x_1(x_2+x'_1+x'_2) -x_2(x'_1+x'_2)-x'_1x'_2\,.\nn
\en
}
In the second case we have to put $x'_{1,2}=x_{1,2}$ in these equations, whereas  in the third case we have
\bq\label{g2-x3}
\begin{array}{l}
 x''_1+x''_2=-x_1-x_2-x'_1+\dfrac{b_2^2- a_4}{a_5},\\
\\
x''_1x''_2=\dfrac{ a_3-2b_1b_2}{ a_5}-(x_1+x_2+x'_1)(x''_1+x''_2)-x_1x_2-x_1x'_1-x_2x'_1.\\
\end{array}
\eq
Four coefficients $b_3,b_2,b_1$ and $b_0$ of  polynomial $\mathcal P(x)$ (\ref{eq-p}) are calculated  by solving  four algebraic equations:
\bq\label{eq-b}
\begin{array}{cl}
1.\quad &y_{1,2}=\mathcal P(x_{1,2})\,,\qquad y'_{1,2}=\mathcal P(x'_{1,2})\,;\\
\\
2.\quad& y_{1,2}=\mathcal P(x_{1,2})\,,\qquad \left.\dfrac{d P(x)}{dx}\right|_{x=x_{1,2}}=\left.\dfrac{d \sqrt{f(x)}}{dx}\right|_{x=x_{1,2}}\equiv \dfrac{1}{2y_{1,2}}\, \partial f(x_{1,2})\,,
\\
\\
3.\quad &y_{1,2}=\mathcal P(x_{1,2})\,,\qquad y'_{1}=\mathcal P(x'_{1})\,,\qquad b_3=0\,,\\
\end{array}
\eq
where $\partial f(x)$ is derivative of $f(x)$ (\ref{eq-c}) by $x$. Substituting  coefficients $b_{k}$ into (\ref{g2-x12}-\ref{g2-x3}) one gets abscissas $\tilde{u}_{1,2}$, whereas the corresponding ordinates $y''_{1,2}$  are equal to
\bq\label{g2-y}
y''_{3,4}=-\mathcal P(x''_{1,2})\,,
\eq
where polynomial $\mathcal P(x)$ is  given by
\bq\label{ab-p}
\begin{array}{lcl}
1.\quad \mathcal  P(x)
&=&\dfrac{(x-x'_2) (x-x'_1) (x-x_2 ) y_1}{(x_1-x'_1 ) (x_1-x'_2) (x_1-x_2)}
+
\dfrac{(x-x'_2) (x-x'_1) (x-x_1 ) y_2}{(x_2-x'_1 ) (x_2-x'_2) (x_1-x_2)}\\
\\
&+&
\dfrac{(x-x'_2) (x-x_2 ) (x-x_1 ) y'_1}{(x'_1-x_1 ) (x'_1-x_2 ) (x'_2-x'_1)}
+
\dfrac{(x-x'_1) (x-x_2 ) (x-x_1 ) y'_2}{ (x'_2-x_1 ) (x'_2-x_2 )(x'_2-x'_1)}\,,\\
\\
2.\quad\mathcal  P(x)&=&
\dfrac{(x-x_2)^2 (2x-3 x_1+x_2) y_1}{(x_2-x_1)^3}
+
\dfrac{(x-x_1)^2 (2x+x_1-3 x_2) y_2}{(x_1-x_2)^3}\\
\\
&+&\dfrac{(x-x_2)^2 (x-x_1) \partial f(x_1)}{2(x_1-x_2)^2 y_1}+
\dfrac{(x-x_1)^2 (x-x_2) \partial f(x_2)}{2(x_1-x_2)^2 y_2}\,,\\
\\
3.\quad
\mathcal P(x)&=&
\dfrac{y_1(x-x_2)(x-x'_1)}{(x_1-x_2)(x_1-x'_1)}+\dfrac{y_2(x-x_1)(x-x'_1)}{(x_2-x_1)(x_2-x'_1)}
+\dfrac{y'_1(x-x_1)(x-x_2)}{(x'_1-x_1)(x'_1-x_2)}\,.\\
\end{array}
\eq
In (\ref{g2-y}) we also made a reduction, which coincides with inversion $(x,y)\to (x,-y)$.

Similar formulae for the hyperelliptic curves  $X$ with $h(x)\neq 0$ in (\ref{h-curve}) are also well known, see \cite{cost12} and references within.

\section{Examples of auto B\"{a}cklund transformations}

\subsection{Lagrange top}
 Let us consider rotation of a rigid body around a fixed point in a  homogeneous gravity  field in the Lagrange case.
 In terms of the Euler angles  $\psi,\theta,\phi$, and momenta $p_\psi,p_\theta,p_\phi$
 the Hamilton function looks like
 \[
 H=p_\theta^2+
 \dfrac{p^2_\phi +2\cos\theta\,p_\phi p_\psi+p^2_\psi}{\sin^2\theta}+ \cos\theta\,.
 \]
 It  is invariant with respect to canonical transformations   of valence $c$
 \bq\label{lag-can1}
  \psi\to \tilde{\psi}=c\psi+a\,,\qquad \phi\to\tilde{\phi}= c\phi+b\,,\qquad \tilde{p}_\psi=p_\psi\,,\qquad\tilde{p}_\phi=p_\phi.
  \eq
For  canonical transformation  $(q,p)\to(\tilde{q},\tilde{p})$ of valence $c$ the Jacobi matrix of transformation
\[
V=\left(
    \begin{array}{cc}
      \dfrac{\partial \tilde{q}}{\partial q} &  \dfrac{\partial \tilde{q}}{\partial p}  \\ \\
       \dfrac{\partial \tilde{p}}{\partial q}  &  \dfrac{\partial \tilde{p}}{\partial p}  \\
    \end{array}
  \right)
\]
 is a generalized symplectic matrix of valence $c$
\[
V^\top  \Omega V=c\Omega \,,\qquad \Omega=\left(\begin{array}{cc}
            0 & Id \\
            -Id & 0 \\
          \end{array}
        \right)\,,
\]
see details in \cite{gant}.

Canonical transformation (\ref{lag-can1}) is a symmetry and, according to Noether's theorem, this symmetry is related to integrals of motion
\bq\label{lag-sep-triv}
A=p_\psi\,,\qquad B=p_\phi.
\eq
We can also use this canonical transformation for construction of integrable discretization
\[ \psi_k=c_k\psi_{k-1}+a_k\,,\qquad \phi_{k}=c_k\psi_{k-1}+b_k\,,\]
for  dynamics of spin $\psi(t)$ and precession $\phi(t)$.

In order to describe similar canonical transformation, symmetry and discretization for nutation  we can use the standard arithmetic on elliptic curves. Indeed, in \cite{lag} Lagrange noted that nutation $u=cos\theta$  is the Weierstrass elliptic function $\wp(t)$ because
 Hamilton-Jacobi equation $H=E$
\bq\label{lagr-curve}
\dfrac{v^2+A^2+2uAB+B^2}{1-u^2}+u=E\,,\quad\mbox{where}\quad v=\sin\theta\, p_\theta\,,
\eq
defines the  elliptic curve. It allows us to identify a tuple of integrals $H, p_\psi,p_\phi$  with a set of points  $P=(\cos\theta,\sin\theta)$, $P_2=(\psi,p_\psi$ and $P_3=(\phi,p_\phi)$ on a direct product  of the plane algebraic curves defined by separation relations
(\ref{lagr-curve}) and (\ref{lag-sep-triv}).

For  elliptic curve
\[
 X:\qquad y^2=f(x)\,,\qquad f(x)= a_3x^3+ a_2x^2+a_1x+a_0
 \]
the Jacobian of $X$ is isomorphic to the curve itself.  So, let us consider points $P=(x,y)$, $P'=(x',y')$ and $P''=(x'',y'')$ on $X$
so that
 \[\mbox{Case 1:}\quad P+P'=P'',\qquad \mbox{Case 2:}\quad [2]P=P''\,,\qquad  \mbox{Case 3:}\quad  [3]P=P''\,.\]
According to \cite{hand06} coordinates of $P''$ are equal to
\bq\label{ell-xy}
\begin{array}{lll}
1.\quad &x''=-x-x'-\dfrac{a_2}{a_3}+\dfrac{1}{a_3}\left(\dfrac{y-y'}{x-x'}\right)^2\,,\quad
&y''=-\dfrac{x''-x'}{x-x'}y+\dfrac{x''-x}{x'-x}y'\,,
\\ \\
2.\quad
&x''=-2x-\dfrac{a_2}{a_3}+\dfrac{1}{a_3}\left(\dfrac{3a_3x^2+2a_2x+a_1}{2y}\right)^2\,,\quad
&y''=-y-\dfrac{(x''-x)(3a_3x^2+2a_2x+a_1)}{2y}\,,
\\ \\
3.\quad
&x''=-3x-\dfrac{2b_1}{b_2}-\dfrac{a_3}{b_2^2}\,,\quad
&y''=b_2{x''}^2+b_1x''+b_0\,,
 \end{array}
 \eq
 where $b_k$ are coefficients of the polynomial
 \[
 \mathcal P(z)=b_2z^2+b_1z+b_0=
 y+\dfrac{(z-x)(3a_3xz+a_2(z+x)+a_1)}{2y}-\dfrac{(z-x)^2(3a_3x^2+2a_2x+a_1)^2}{8y^3}
 \]
Substituting  equation (\ref{lagr-curve}) for the Lagrange curve into these standard equations we obtain
 \ben\label{pq-lag}
1.\quad &\tilde{u}=-u-\lambda+H+\left(\dfrac{v-\mu}{u-\lambda}\right)^2\,,\quad
&\tilde{v}=-\dfrac{\tilde{u}-\lambda}{u-\lambda}\,v+\dfrac{\tilde{u}-u}{\lambda-u}\,\mu\,,
 \\
 \nn\\
2.\quad
&\tilde{u}=-2u+H-\dfrac{(2AB+2Hu-3u^2+1)^2}{4v^2}\,,\quad
&\tilde{v}=-v-\dfrac{(\tilde{u}-u)(2AB+2Hu-3u^2+1)}{2v}\,.
 \nn
 \en
 Here we  put $(x,y)=(u,v)$, $(x'',y'')=(\tilde{u},\tilde{v})$ and $(x',y')=(\lambda,\mu)$ in order to distinguish
 points on the $(x,y)$-plane and  coordinates on the phase space.  In the third case we present polynomial $\mathcal P(z)$ only:
 \[
 \mathcal P(z)=v-\dfrac{(z-u)(2AB+H(z+u)-3zu+1)}{2v}-\dfrac{(z-u)^2(2AB+2Hu-3u^2+1)^2}{8v^3}\,.
 \]

 \begin{prop}
Equations (\ref{pq-lag}) determine canonical transformations $(\theta,p_\theta)\to (\tilde{\theta},\tilde{p}_\theta)$ of valencies one, two and three, respectively.  These canonical transformations preserve the form of Hamilton-Jacobi equation $H=E$, i.e.
they are auto B\"{a}cklund transformations for the Lagrange top.
\end{prop}
The proof is a straightforward calculation in which we have taken into account that $\mu=\sqrt{f(\lambda)}$ is  a function on $\theta$,  $p_\theta$ and parameter $\lambda$.

\subsection{H\'{e}non-Heiles system}
Let us  take H\'{e}non-Heiles system with Hamiltonians
\begin{equation}\label{hh-1}
H_1=\frac{p_1^2+p_2^2}{4}-4aq_2(q_1^2+2q_2^2)\,,\qquad
H_2=\frac{p_1(q_1p_2-q_2p_1)}{2}-aq_1^2(q_1^2+4q_2^2)\,
\end{equation}
separable in parabolic coordinates on the plane
 \begin{equation}\label{u-par}
u_1 = q_2-\sqrt{q_1^2+q_2^2},\qquad u_2 = q_2+\sqrt{q_1^2+q_2^2}\,.
\end{equation}
Standard momenta associated with parabolic coordinates $u_{1,2}$ are equal to
\[
p_{u_1}=\frac{p_2}{2}-\frac{p_1(q_2+\sqrt{q_1^2+q_2^2})}{2q_1}\,,\qquad
p_{u_2}=\frac{p_2}{2}-\frac{p_1(q_2-\sqrt{q_1^2+q_2^2})}{2q_1}\,.
\]
To describe evolution of $u_ {1,2} $ with respect to $H_ {1,2} $ we use the canonical Poisson bracket
\bq\label{poi1}
\{q_i,p_{j}\}=\delta_{ij}\,,\quad \{q_1,q_2\}=\{p_{1},p_{2}\}=0\,,\qquad
\{u_i,p_{u_j}\}=\delta_{ij}\,,\quad \{u_1,u_2\}=\{p_{u_1},p_{u_2}\}=0\,.
\eq
and expressions for $H_{1,2}$
\bq\label{hh-ham-u}
H_1=\frac{\scriptstyle p_{u_1}^2u_1-p_{u_2}^2u_2}{\scriptstyle u_1-u_2}-a(u_1+u_2)(u_1^2+u_2^2)\,,\quad
H_2=\frac{\scriptstyle u_1u_2(p_{u_1}^2-p_{u_2}^2)}{\scriptstyle u_2-u_1}+au_1u_2(u_1^2+u_1u_2+u_2^2)
\eq
 to obtain
\begin{equation}\label{hh-eq1}
\frac{du_1}{dt_1}=\{u_1,H_1\}=\frac{2p_{u_1}u_1}{u_1-u_2}\,,\qquad
\frac{du_2}{dt_1}=\{u_2,H_1\}=\frac{2p_{u_2}u_2}{u_2-u_1}
\end{equation}
and
\begin{equation}\label{hh-eq2}
\frac{du_1}{dt_2}=\{u_1,H_2\}=\frac{2u_1u_2p_{u_1}}{u_2-u_1}\,,\qquad
\frac{du_2}{dt_2}=\{u_2,H_2\}=\frac{2u_1u_2p_{u_2}}{u_1-u_2}\,.
\end{equation}
Using Hamilton-Jacobi equations $H_{1,2}=\alpha_{1,2}$ we can prove that these variables satisfy to the following separated relations
\begin{equation}\label{hh-sep}
\bigl(u_ip_{u_i}\bigr)^2=u_i(au_i^4+\alpha_1u_i+\alpha_2)\,,\qquad i=1,2.
\end{equation}
Expressions (\ref{hh-eq1}-\ref{hh-eq2} ) and (\ref{hh-sep}) yield standard Abel quadratures
\bq\label{hh-ab-q1}
\frac{du_1}{\sqrt{f(u_1)}}+\frac{du_2}{\sqrt{f(u_2)}}=2dt_2\,,\qquad \frac{u_1du_1}{\sqrt{f(u_1)}}+\frac{u_2du_2}{\sqrt{f(u_2)}}=2dt_1,
\eq
on   hyperelliptic curve $ X$ of genus two  defined by equation
\bq\label{hh-eq-c}
X:\quad y^2=f(x)\,,\qquad f(x)=x(ax^4+\alpha_1x+\alpha_2)\,.
\eq

Suppose that transformation of variables
\bq\label{st-b-hh} (q_{1},q_2,p_1,p_{2})\to (\tilde{q}_{1},\tilde{q}_2,\tilde{p}_{1},\tilde{p}_{2})
\eq
preserves Hamilton equations (\ref{hh-eq1}-\ref{hh-eq2}) and the form of Hamiltonians (\ref{hh-ham-u}). It means that new parabolic coordinates  $\tilde{u}_{1,2}=\tilde{q}_2\pm \sqrt{\tilde{q}_1^2+\tilde{q}_2^2}$ satisfy to the same equations
\bq\label{hh-ab-q2}
\frac{d\tilde{u}_1}{\sqrt{f(\tilde{u}_1)}}+\frac{d\tilde{u}_2}{\sqrt{f(\tilde{u}_2)}}=2dt_2\,,\qquad \frac{\tilde{u}_1d\tilde{u}_1}{\sqrt{f(\tilde{u}_1)}}+\frac{\tilde{u}_2d\tilde{u}_2}{\sqrt{f(\tilde{u}_2)}}=2dt_1\,.
\eq
Subtracting (\ref{hh-ab-q2}) from (\ref{hh-ab-q1}) one gets Abel differential equations
\begin{equation}\label{ab-eq-g2}
\begin{array}{c}
\omega_1(x_1,y_1)+\omega_1(x_2,y_2)+\omega_1(x''_1,y''_1)+\omega_1(x''_2,y''_2)=0\,,\\ \\
\omega_2(x_1,y_1)+\omega_2(x_2,y_2)+\omega_2(x''_2,y''_1)+\omega_2(x''_2,y''_2)=0\,,
\end{array}
\end{equation}
where
 \[
 x_{1,2}=u_{1,2},\quad y_{1,2}=u_{1,2}p_{u_{1,2}}\,,\qquad
x''_{1,2}=\tilde{u}_{1,2},\quad y''_{1,2}=-\tilde{u}_{1,2}\tilde{p}_{u_{1,2}}
\]
and $\omega_{1,2}$ form a base of holomorphic differentials on hyperelliptic curve $X$ of genus $g=2$
\[
\omega_1(x,y)=\frac{dx}{y}\,,\qquad \omega_2(x,y)=\frac{xdx}{y}\,.
\]
Solutions of the Abel equations form so-called intersection divisor of two plane curves  $X$ and $Y$ \cite{ab,kl05}, that allows us directly  apply cryptographic protocols based on  arithmetic of divisor to construction of canonical transformation (\ref{st-b-hh}) on the phase space.

Let us suppose that generic points  $P_1=(x_{1},y_{1})$ and $P_2=(x_{2},y_{2})$ form divisor $D$ (message) in (\ref{cases}), whereas points  $P''_{1,2}=(x''_{1,2},y''_{1,2})$  belong to support of resulting divisor $D''$ (cryptogram).  Canonical transformations (\ref{st-b-hh}) associated with arithmetic operations  (\ref{cases}) are completely defined by coefficients $b_k$ of polynomial $\mathcal P$ (\ref{ab-p}). In the first case, these coefficients  $b_k$ are defined by
  \ben
  Ab_0&=&q_1x_1'x'_2(x'_1-x'_2)\Bigl(p_1\bigl(q_1^2+4q_2^2-2q_2(x'_1+x'_2)+x'_1x'_2\bigr)+
q_1p_2(x'_1+x'_2+2q_2)\Bigr)
\nn\\
\nn\\
&-&2q_1^2x'_2(q_1^2+2q_2x'_2-{x'_2}^2)y'_1+2q_1^2x'_1(q_1^2+2q_2x'_1-{x'_2}^2)y'_2;
\nn\\
\nn\\
   Ab_1&=&(x'_2-x'_1)\left(
q_1p_1\bigl((q_1^2+4 q_2^2)(x'_1+x'_2)-2q_2({x'_1}^2+x'_1x'_2+ {x'_2}^2) \bigr)\right.\nn\\
&&\qquad\qquad
-\left.p_2\bigl (2 q_1^2 q_2(x'_1+x'_2)-q_1^2({x'_1}^2+x'_1x'_2+ {x'_2}^2)+{x'_1}^2{ x'_2}^2\bigr)\right)\nn\\
&+&2 (q_1^2-2 q_2 x'_2) (q_1^2+2 q_2x'_2-{x'_2}^2) y'_1-2 (q_1^2-2 q_2 x'_1) (q_1^2+2 q_2x'_1-{x'_1}^2) y'_2\,,
\nn\\
\nn\\
Ab_2&=& (x'_1-x'_2)\left(q_1p_1(q_1^2+4 q_2^2-{x'_1}^2-x'_1x'_2-{x'_2}^2) - \bigl(2q_1^2q_2+(x'_1+x'_2)x'_1x'_2\bigr) p_2 \right) \nn\\
  \nn\\
  &+&2(2 q_2+x'_2) (q_1^2+2 q_2 x'_2-{x'_2}^2)y'_1-2 (2 q_2+x'_1) (q_1^2+2 q_2 x'_1-{x'_1}^2) y'_2\,,\nn\\
  \nn\\
 Ab_3&=&(x'_2-x'_1)\left(q_1p_1(2q_2-x'_1-x'_2)-(q_1^2+x'_1x'_2)p_2\right)-2(q_1^2+2q_2x'_2-{x'_2}^2)y'_1\nn\\
 \nn\\
 &-&2(q_1^2+2q_2x'_1-{x'_1}^2)y'_2\,,
 \nn
\en
where
\[
A=2(x'_1-x'_2)(q_1^2+2q_2x'_1-{x'_2}^2)(q_1^2+2q_2x'_2-{x'_2}^2).
\]
Substituting these coefficients into the following expressions one gets explicit formulae for the new variables
\bq\label{pq-case1}
\begin{array}{l}
\tilde{q}_2= -q_2-\dfrac{x'_1+x'_2}{2}-\dfrac{b_2}{b_3}+\dfrac{a}{2b_{3}^2}\,,
\\
\\
\tilde{q}_1^2 =-q_1^2+2\tilde{q}_2(2q_2+x'_1+x'_2)+2q_2(x'_1+x'_2)+x'_1x'_2-\dfrac{2b_1}{b_3}-\dfrac{b_2^2}{b_3^2}\,,\\
\\
\tilde{p}_1= -\dfrac{2b_0}{\tilde{q}_1}-4\tilde{q}_1\tilde{q_2}b_3-2\tilde{q}_1b_2\,,\quad
\tilde{p}_2=-2(\tilde{q}_1^2+4\tilde{q}_2^2)b_3-4\tilde{q}_2b_2-2b_1\,.
\end{array}
\eq
In  Case 2 we have
\ben
Bb_0&=& -8aq_1^4(2p_1q_1q_2-p2q_1^2-2p_2q_2^2)+q_1^2p_1(p_1^2q_1+2p_1p_2q_2-2p2^2q_1)\,,\nn\\
\nn\\
Bb_1&=&-8aq_1^2(p_1q_1^3+10p_1q_1q_2^2-4p_2q_1^2q_2-4p_2q_2^3)-2p_1^3q_1q_2+3p_1^2q_1^2p_2-4p_1^2p_2q_2^2\nn\\
\nn\\
&+&4p_1p_2^2q_1q_2-2p_2^3q_1^2\,,\nn\\
\nn\\
Bb_2&=&-8aq_1(12p_1q_2^3+p_2q_1^3-2p_2q_1q_2^2)+p_1^2(p_1q_1+4p_2q_2)\,,\nn\\
\nn\\
Bb_3&=&8aq_1(p_1q_1^2+6p_1q_2^2-2p_2q_1q_2)-p_2p_1^2\,,\qquad B = 4q_1(p_1^2q_1+2p_1p_2q_2-p_2^2q_1)\nn
\en
so that
 \ben
  \tilde{q}_2&=&- 2q_2+\dfrac{1}{\bigl(8q_1(p_1q_1^2+6p_1q_2^2-2p_2q_1q_2)a-p_2p_1^2\bigr)^2}\nn\\
 \nn\\
 &\times&\Bigl(
64 q_1^2 (12 p_1 q_2^3+p_2 q_1^3-2 p_2 q_1 q_2^2) (p_1 q_1^2+6 p_1 q_2^2-2 p_2 q_1 q_2) a^2+p_1^4 p_2 (p_1 q_1+4 p_2 q_2)\Bigr.
\nn\\
\nn\\
&&\Bigl.-8a q_1 (6 p_1^4 q_1 q_2^2-2 p_1^3 p_2 q_1^2 q_2+36 p_1^3 p_2 q_2^3+3 p_1^2 p_2^2 q_1^3-14 p_1^2 p_2^2 q_1 q_2^2+4 p_1 p_2^3 q_1^2 q_2-p_2^4 q_1^3)\Bigr)\,,
\nn\\
\nn\\
\tilde{q}_1^2&=&-\left(\dfrac{
8aq_1^2(2p_1q_1q_2-p_2q_1^2-2p_2q_2^2)-p_1(p_1^2q_1+2p_1p_2q_2-2p_2^2q_1
}{8aq_1(p_1q_1^2+6p_1q_2^2-2p_2q_1q_2)-p_2p_1^2}\right)^2\,,\label{pq-case2}\\
\nn\\
\tilde{q}_1\tilde{p}_1&=& -2b_0-2\tilde{q}^2_1(b_2+2\tilde{q_2}b_3)\,,\quad
\tilde{p}_2=-2(\tilde{q}_1^2+4\tilde{q}_2^2)b_3-4\tilde{q}_2b_2-2b_1\,.\nn
\en
In Case 3 coefficients are equal to
  \bq\label{b-case3}
 \begin{array}{l}
 b_0=\dfrac{ (x'_1(2 p_1 q_2 x'_1-p_2 q_1)p_1 {x'_1}^2+2 q_1 y'_1) q_1}{2(q_1^2+2 q_2 x'_1-{x'_1}^2)}
 \\
 b_1= -\dfrac{2 p_1 q_1 q_2-p_2(q_1^2-{x'_1}^2)-4 q_2 y'_1}{2(q_1^2+2 q_2 x'_1-{x'_1}^2)}\,,
 \qquad
 b_2=\dfrac{p_1 q_1+p_2 x'_1-2 y'_1}{2(q_1^2+2 q_2 x'_1-{x'_1}^2)}\,,
 \end{array}
 \eq
 so that
 \ben
 \tilde{q}_2&=&-q_2-\dfrac{x'_1}{2}+\dfrac{b_2^2}{2a}=-q_2-\dfrac{x'_1}{2}+\dfrac{(p_1q_1+p_2 x'_1-2y'_1)^2}{8a(q_1^2+2q_2 x'_1-{x'_1}^2)^2}\,,\nn\\
 \nn\\
  \tilde{q}_1^2&=&-q_1^2+2\tilde{q}_2(2q_2+x'_1)+2q_2x'_1+\dfrac{2b_1b_2}{a}\label{pq-case3}\\
  \nn\\
  &=&-q_1^2+2\tilde{q}_2(2q_2+x'_1)+2q_2x'_1
 -\frac{(p_1q_1+p_2x'_1-2y'_1)(2p_1q_1q_2-p_2q_1^2+p_2{x'_1}^2-4q_2y'_1)}{2a(q_1^2+2q_2x'_1-{x'_1}^2)^2}\,,\nn\\
 \nn\\
 \tilde{p}_2&=& -4\tilde{q}_2b_2-2b_1=-p_2
 -\dfrac{2(q_2-\tilde{q}_2)(2y'_1-p_1q_1-p_2x'_1)}{q_1^2+2q_2x'_1-{x'_1}^2}\,,\nn\\
 \nn\\
 \tilde{q}_1 \tilde{p}_1&=&-2b_0-2q^2_1b_2= -q_2p_2
 -\dfrac{(q_1^2-\tilde{q}^2_2)(2y'_1-p_1q_1-p_2x'_1)}{q_1^2+2q_2x'_1-{x'_1}^2}\,.\nn
  \en
These  explicit formulae for $\tilde{q}_{1,2}$ and $\tilde{p}_{1,2}$ can be easily obtained using any  modern  computer algebra system.

We present  these bulky expressions here only  so that any reader can verify that these transformations $(q,p)\to(\tilde{q},\tilde{p})$ are different, i.e. cannot be obtained from each other for special values of parameters, and that these transformations have the following properties.
\begin{prop}
Equations (\ref{pq-case1},\ref{pq-case2}) and (\ref{pq-case3})   determine canonical transformations (\ref{st-b-hh}) on $T^*\mathbb R^2$ of valencies one and two for which original Poisson bracket (\ref{poi1}) has the following form in new variables
\[\begin{array}{cl}
1,3.\qquad&\{\tilde{q}_i,\tilde{p}_{j}\}=\phantom{2}\delta_{i,j}\,,\qquad \{\tilde{q}_1,\tilde{q}_2\}=\{\tilde{p}_{1},\tilde{p}_{2}\}=0\,,
\\
\\
 2.\quad&
\{\tilde{q}_i,\tilde{p}_{j}\}=2\delta_{i,j}\,,\qquad \{\tilde{q}_1,\tilde{q}_2\}=\{\tilde{p}_{1},\tilde{p}_{2}\}=0\,,
\end{array}
\]
respectively.  These canonical transformations preserve the form of integrals of motion  (\ref{hh-1}), i.e. they are auto B\"{a}cklund transformations in the Toda-Wadati sense \cite{toda75}.
\end{prop}
The proof is a straightforward calculation.

 The fact most interesting to us is that obtained cryptograms are new canonical variables on the original phase space, which can be used  for construction of new integrable systems in the framework of the Jacobi method.

 For instance, let us suppose that  divisor $D'$ (secret key) consists of  ramification point $P_0=(0,0)$ and point $P_\infty$. In this case cryptogram, i.e canonical variables $\tilde{u}_{1,2},\tilde{p}_{u_{1,2}}$ are defined by equations (\ref{g2-x3}) and (\ref{g2-y}) for $x''_k=\tilde{u}_k$ and $y''_k=\tilde{u}_k\tilde{p}_{u_k}$:
\[
\begin{array}{l}
 \tilde{u}_1+\tilde{u_2}=-u_1-u_2+\dfrac{b_2^2}{a},\quad
\tilde{u}_1\tilde{u}_2=-\dfrac{2b_1b_2}{ a}-(u_1+u_2)(\tilde{u}_1+\tilde{u}_2)-u_1u_2,\\
\\
\tilde{p}_{u_1}=-\dfrac{b_2\tilde{u}_1^2+b_1\tilde{u}_1+b_0}{\tilde{u}_1}\,,\qquad \tilde{p}_{u_2}=-\dfrac{b_2\tilde{u}_2^2+b_1\tilde{u}_2+b_0}{\tilde{u}_2}\,,
\end{array}
\]
where coefficients $b_k$ are given by (\ref{b-case3}). Substituting  $y=\tilde{p}_{1,2}$ and $x=\tilde{u}_{1,2}$ into the separated relation
\[
\tilde{X}:\quad
(y^2-ax^3-\tilde{H}_1-\tilde{H}_2)(y^2-ax^3-\tilde{H}_1+\tilde{H}_2)+abx+acy=0\,,
\]
which defines genus three hyperelliptic curve $\tilde{X}$, and solving the resulting  equations with respect to $\tilde{H}_{1,2}$, one gets Hamiltonian
\[
\tilde{H}_1=\dfrac{p_1^2}{8}+\dfrac{p_2^2}{4}-aq_2(3q_1^2+8q_2^2)+\dfrac{b}{2q_1^2}-\dfrac{cp_1}{q_1^3}\,.
\]
We can identify this Hamiltonian with well-known second integrable  H\'{e}non-Heiles system with quartic additional integral $H_2$ \cite{ts15a,ts15c}.

According \cite{ts15b,ts17v,ts17c,ts17p,ts17e} we can use these cryptographic protocols in order to get  new integrable systems on the  plane, sphere and ellipsoid with polynomial integrals of motion of sixth, fourth and third order in momenta.

The work was supported by the Russian Science Foundation  (project  15-12-20035).

\end{document}